\documentclass[11pt,twoside,usenatbib]{article}
\usepackage{asp2010}
\pdfoutput=1

\resetcounters

\begin{document}

\title{Ensemble asteroseismology of red-giant stars}
\author{S. Hekker$^1$, R.L. Gilliland$^2$, Sarbani Basu$^3$, J. De Ridder$^4$, W.J. Chaplin$^5$, Y. Elsworth$^5$
\affil{$^1$Astronomical Institute `Anton Pannekoek', University of Amsterdam, Science Park 904, 1098 HX Amsterdam, The Netherlands (E-mail: S.Hekker@uva.nl)}
\affil{$^2$Space Telescope Science Institute, 3700 San Martin Drive, Baltimore, MD 21218, USA}
\affil{$^3$ Department of Astronomy, Yale University, P.O. Box 208101, New Haven CT 06520-8101, USA}
\affil{$^4$Instituut voor Sterrenkunde, K.U. Leuven, Celestijnenlaan 200D, 3001 Leuven, Belgium}
\affil{$^5$ School of Physics and Astronomy, University of Birmingham, Edgbaston, Birmingham B15 2TT, UK}}

\begin{abstract}
The successful launches of the CoRoT and \textit{Kepler} space missions have led to the detections of solar-like oscillations in large samples of red-giant stars. The large numbers of red giants with observed oscillations make it possible to investigate the properties of the sample as a whole: ensemble asteroseismology.

In this article we summarise ensemble asteroseismology results obtained from data released by the \textit{Kepler} Science Team ($\sim$150\,000 field stars) as presented by \citet{hekker2011pub} and for the clusters NGC~6791, NGC~6811 and NGC~6819 \citep{hekker2011clus} and we discuss the importance of such studies. 
\end{abstract}

\section{Introduction}
Data from the CoRoT \citep{baglin2006} and \textit{Kepler} \citep{borucki2009} missions reveal that the majority of stars show intrinsic photometric variability. This variability can be caused by, for instance, stellar activity, such as spots; eclipses in multiple systems consisting of stars or star-planet systems; global intrinsic stellar oscillations or a combination of phenomena. We are interested in stellar oscillations to study the internal structures of stars.

Stellar oscillations can be stochastically excited in turbulent layers, such as the convective zones in the Sun, low-mass main-sequence stars, subgiants and red-giant stars. Indeed in many of these stars, these so-called solar-like oscillations have been observed \citep[see for recent overviews][]{bedding2011,jcd2011,hekker2010}. The properties of solar-like oscillations can be understood in terms of their asymptotic approximation \citep{tassoul1980}, which predicts that acoustic oscillation modes have evenly spaced frequencies with large and small separations between modes of the same degree and consecutive radial orders (large separation, $\Delta \nu$) or modes of different degrees (small separations). The oscillations are centred around the frequency of maximum oscillation power ($\nu_{\rm max}$). The global oscillation parameters ($\nu_{\rm max}$ and $\Delta \nu$) can be used to derive the stellar masses and radii from scaling relations \citep{kjeldsen1995}. Although these masses and radii are not as accurate and precise as can be obtained from grid-based models \citep{gai2011}, which take stellar evolution into account, these are valuable parameters for ensemble studies as discussed here.

As well as the ability to determine fundamental stellar parameters in a relatively straightforward way, a large sample of stars needs to be observable for ensemble studies. All stars in the mass range between roughly 0.8 and 6 solar masses go through a red-giant phase and the stars in this phase are intrinsically bright. Hence, many red giants are observable in the apparent magnitude volume to which both CoRoT and \textit{Kepler} are sensitive. Furthermore, the oscillations in these stars have periods of the order of a few hours to days, which results in frequencies below the Nyquist frequencies of both instruments. Therefore, all the prerequisites for ensemble asteroseismology are available.

In this article we summarise the results on ensemble asteroseismology obtained by \citet{hekker2011clus,hekker2011pub} and discuss the importance of such studies.

\section{Ensembles}
\subsection{Field stars}
The \textit{Kepler} Science Team has released timeseries data for about $\sim$150\,000 stars selected on the basis of allowing planets of 2~R$_{\rm Earth}$ in an orbit as close as 5~R$_{\rm star}$ to be detected if transiting these stars \citep{batalha2010}. Among these stars $\sim$17\,000 are red-giant stars \citep{ciardi2010,hekker2011pub}. In this article data from the first month (run Q1) of science operation have been used together with effective temperatures from ground-based multi-colour photometery available in the $\textit{Kepler}$ Input Catalogue \citep[KIC,][]{brown2011}. 

\subsection{Clusters}
There are four clusters in the \textit{Kepler} field of view: NGC~6791, NGC~6811, NGC~6819 and NGC~6866. 
NGC~6791 is one of the oldest, most massive and most metal-rich open clusters known \citep{origlia2006,carretta2007,anthony2007} and has been studied extensively. The ages proposed for NGC~6791 range from 7 to 12 Gyr \citep[see e.g.,][]{basu2011,grundahl2008}. In addition, there are four independent studies available to determine the metallicity of this cluster, which range from $+0.29 \leq$~[Fe/H]~$\leq +0.45$ \citep{carraro2006,origlia2006, anthony2007,brogaard2011}.  
NGC~6819 is a very rich open cluster with roughly solar metallicity of [Fe/H] = +0.09 $\pm$ 0.03 \citep{bragaglia2001}, an age of about 2.5 Gyr \citep{kalirai2001,kalirai2004} and reddening $E(B-V)$ = 0.15. NGC~6811 is a young, sparse, not particularly well studied cluster. Studies on this cluster tended to focus on membership or variability of stars and no direct metallicity studies are available. Red giants have been observed in these three clusters. The fourth and youngest cluster NGC~6866 has an age of about 0.56 Gyr \citep{frolov2010} and no red giants have been observed in this cluster.

For the clusters we have timeseries data of about two years length at our disposal within the \textit{Kepler} Asteroseismic Science Consortium (KASC). Although these ensembles are much smaller, i.e., 46, 42, and 4 stars for NGC~6791, NGC~6819, NGC~6811 respectively, than the field star sample, the additional information such as common metallicity and age within the clusters allow for an ensemble investigation. The effective temperatures of the cluster stars have been determined using the colour--temperature calibrations by \citet{ramirez2005}. See \citet{hekker2011clus} for more details.

\section{Results}
The analyses of the ensembles described in the previous section led to the following main conclusions:
\subsection{Field stars}
\begin{figure*}
\begin{minipage}{0.5\linewidth}
\centering
\includegraphics[width=\linewidth]{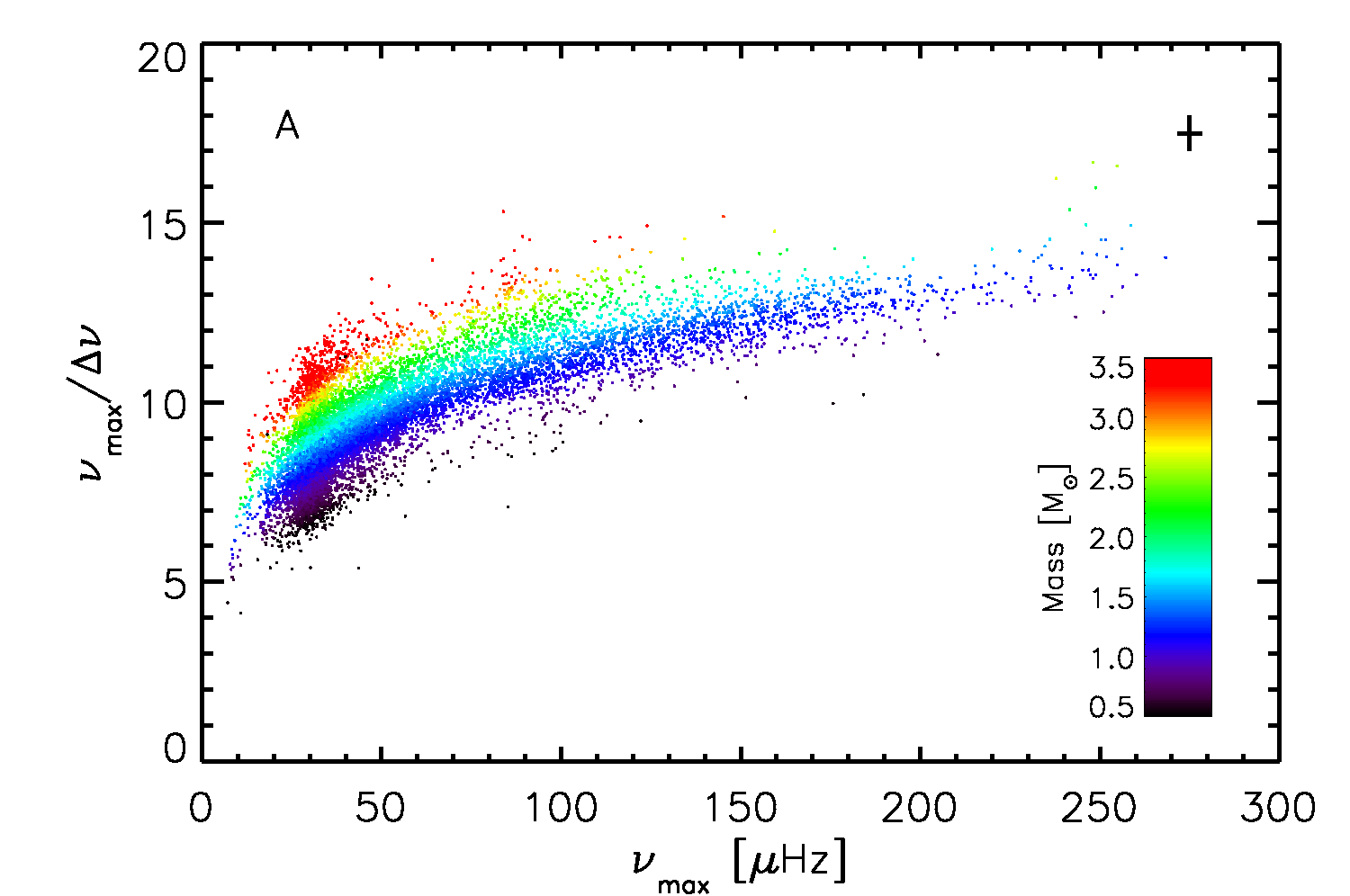}
\end{minipage}
\begin{minipage}{0.5\linewidth}
\centering
\includegraphics[width=\linewidth]{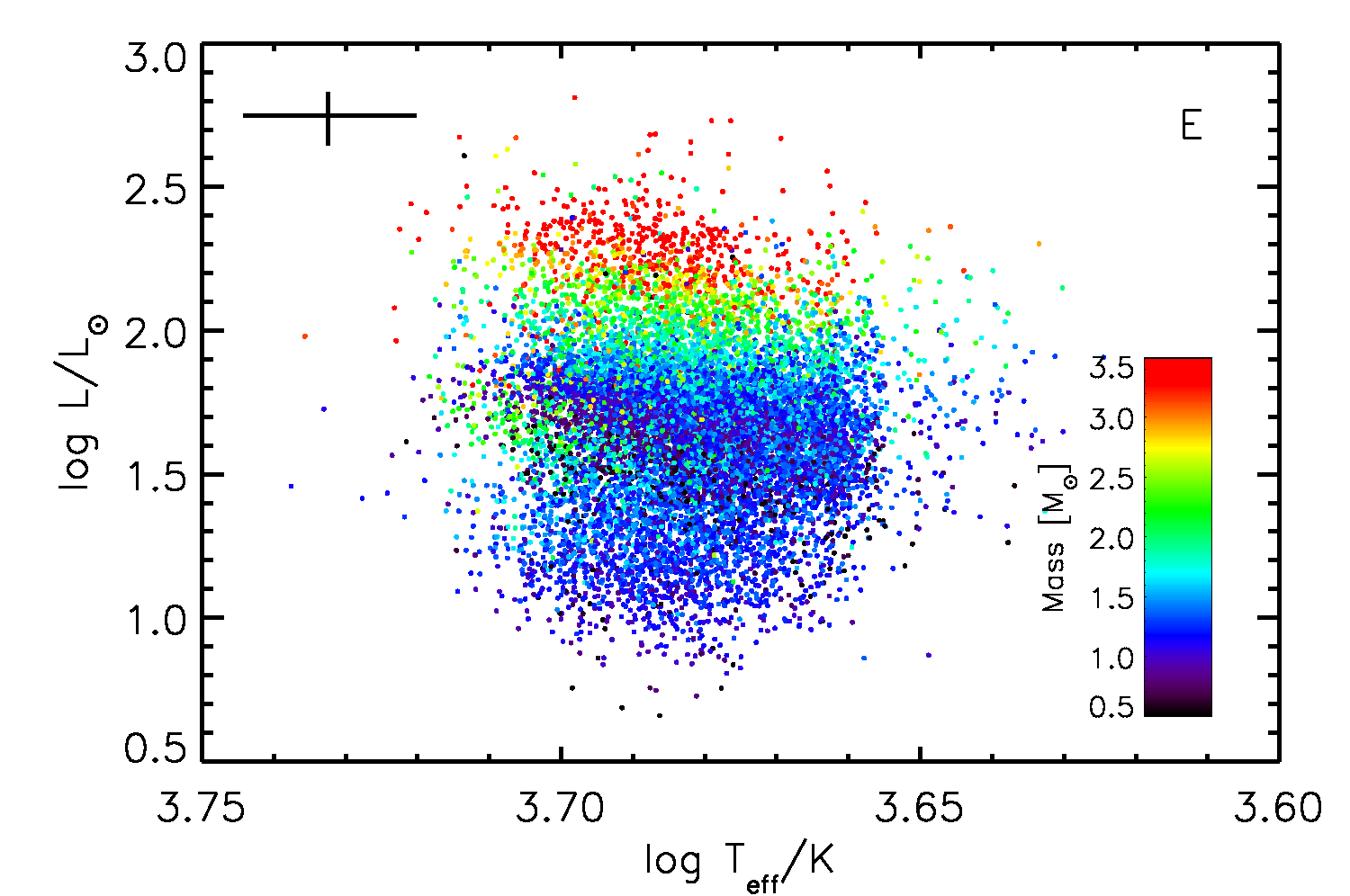}
\end{minipage}
\begin{minipage}{0.5\linewidth}
\centering
\includegraphics[width=\linewidth]{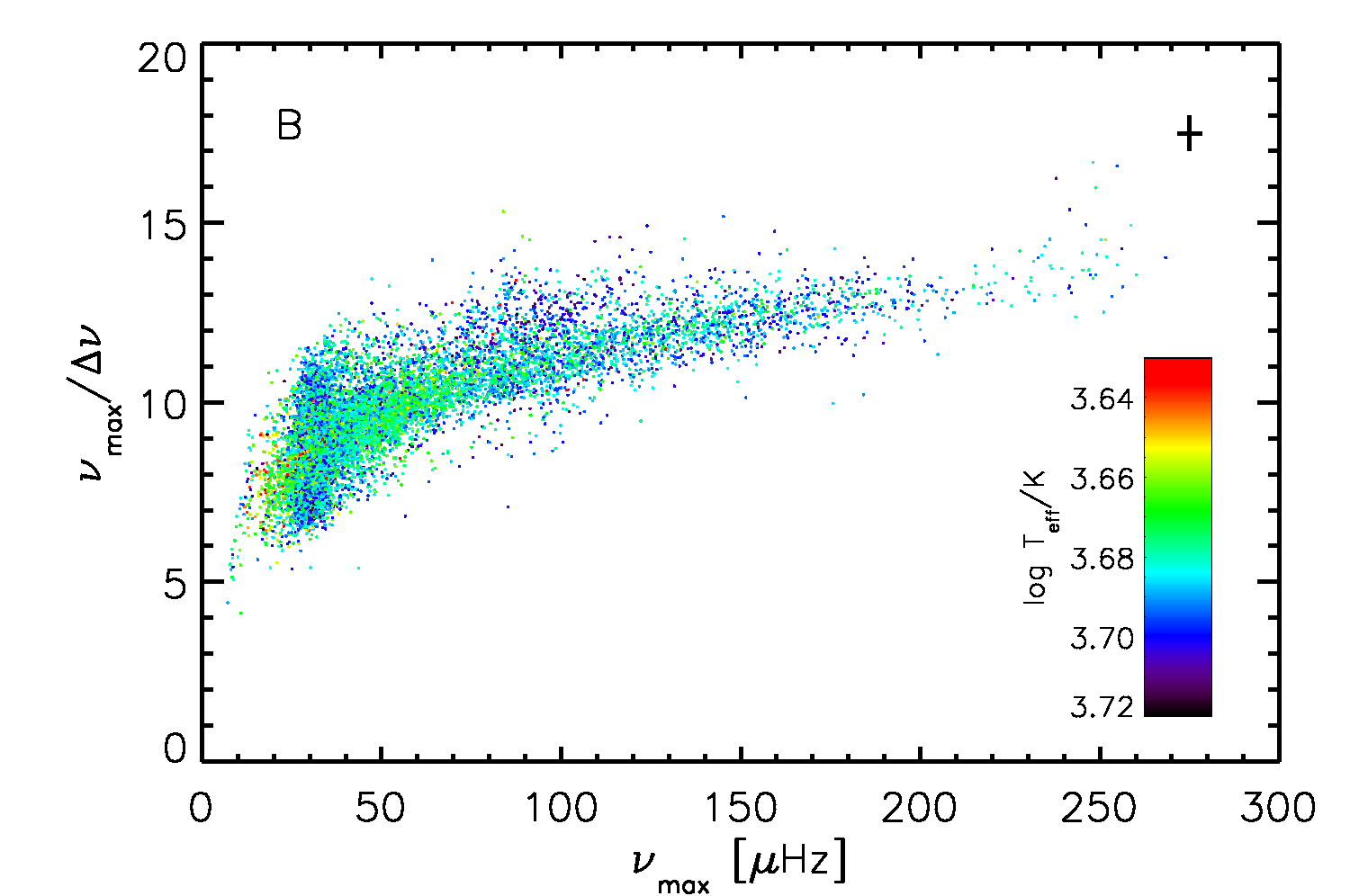}
\end{minipage}
\begin{minipage}{0.5\linewidth}
\centering
\includegraphics[width=\linewidth]{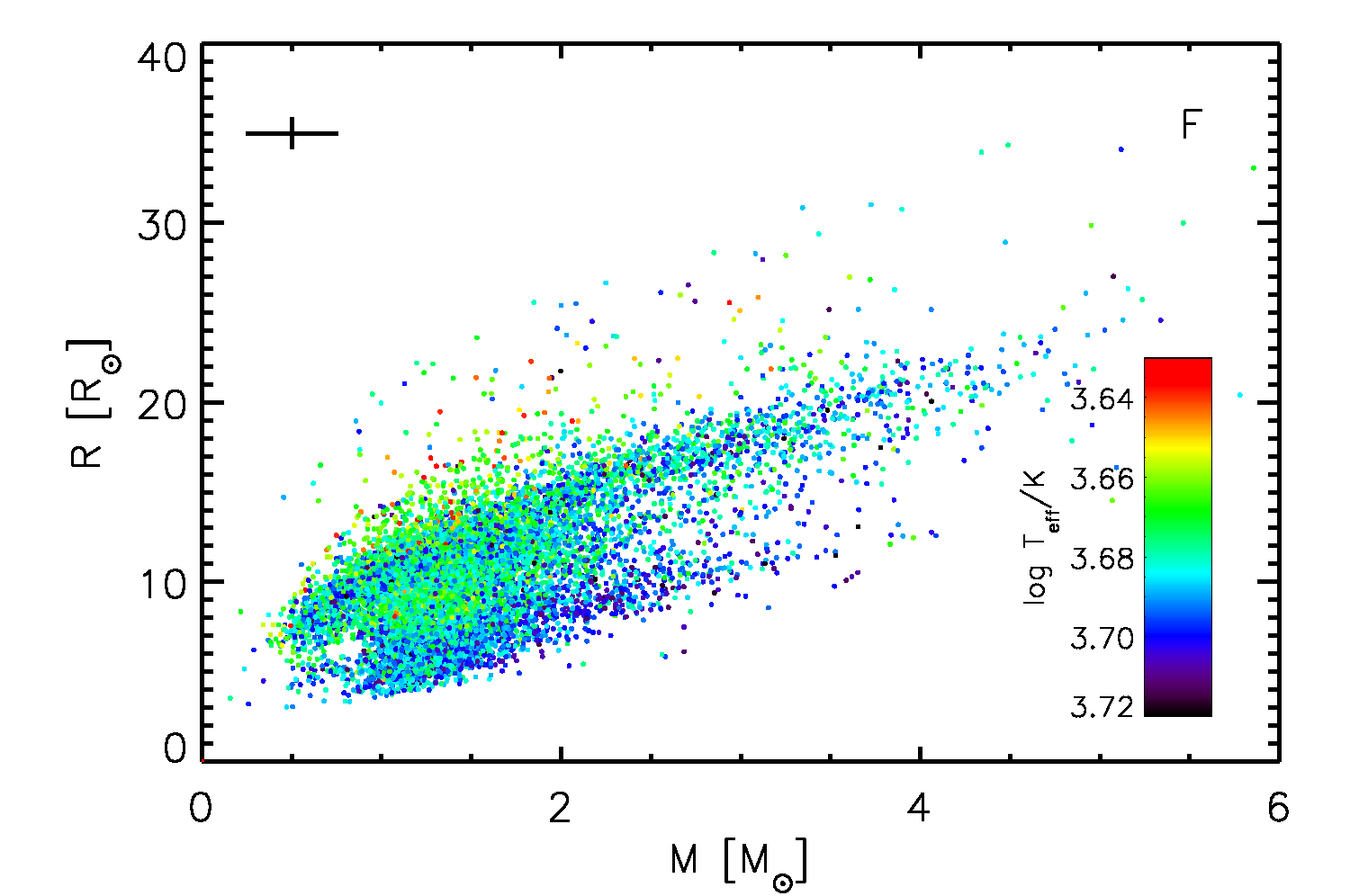}
\end{minipage}
\begin{minipage}{0.5\linewidth}
\centering
\includegraphics[width=\linewidth]{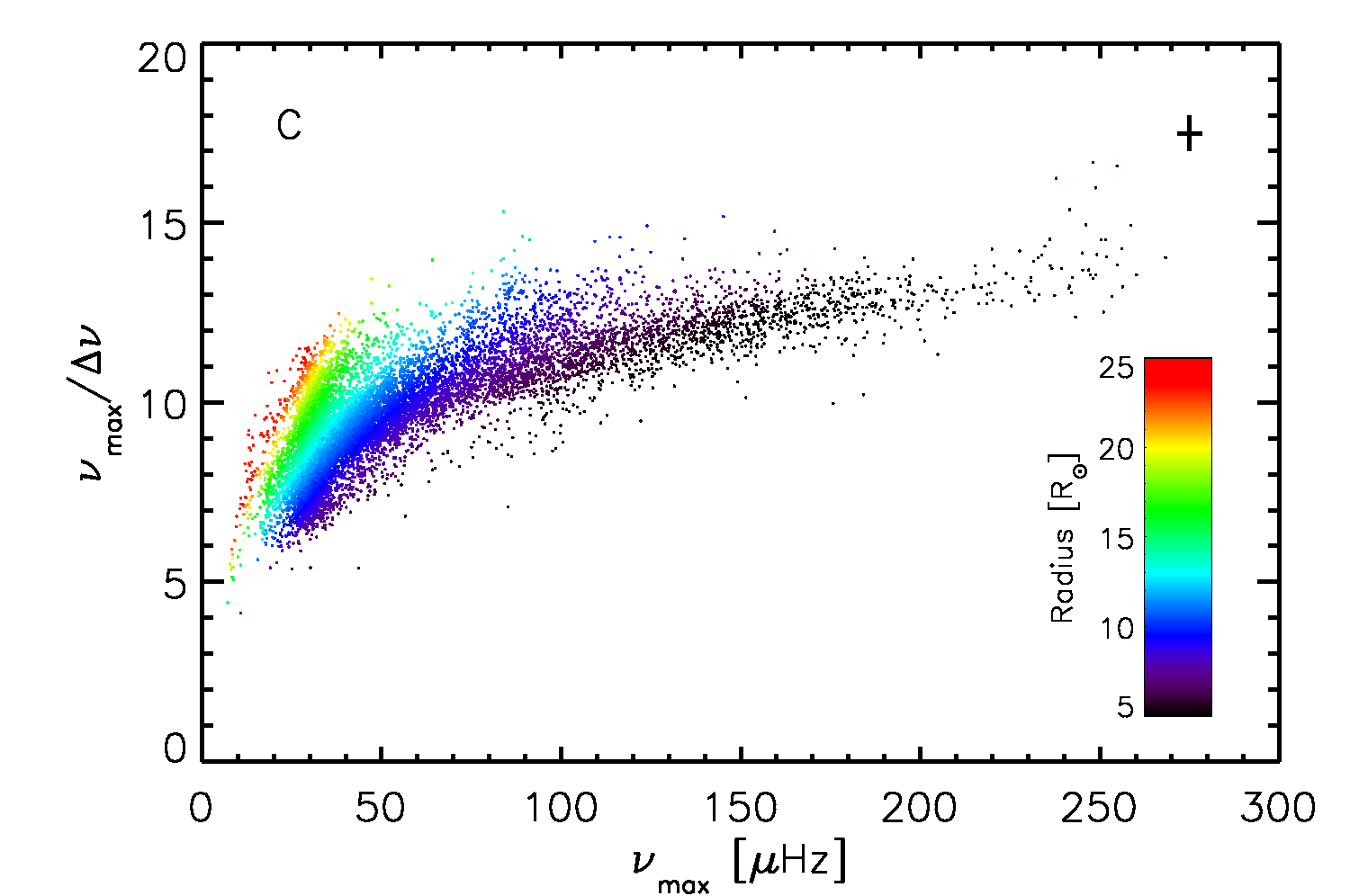}
\end{minipage}
\begin{minipage}{0.5\linewidth}
\centering
\includegraphics[width=\linewidth]{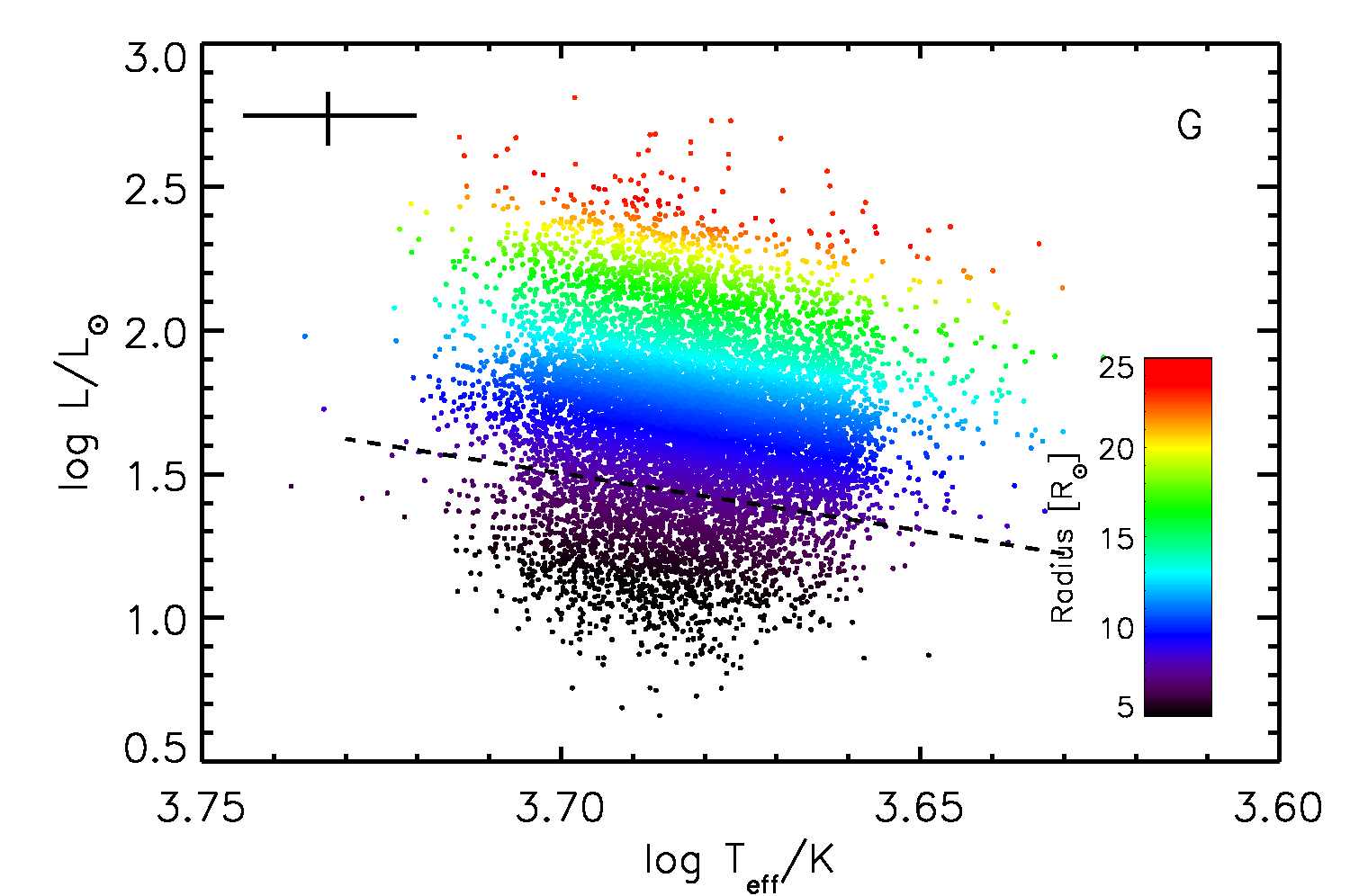}
\end{minipage}
\begin{minipage}{0.5\linewidth}
\centering
\includegraphics[width=\linewidth]{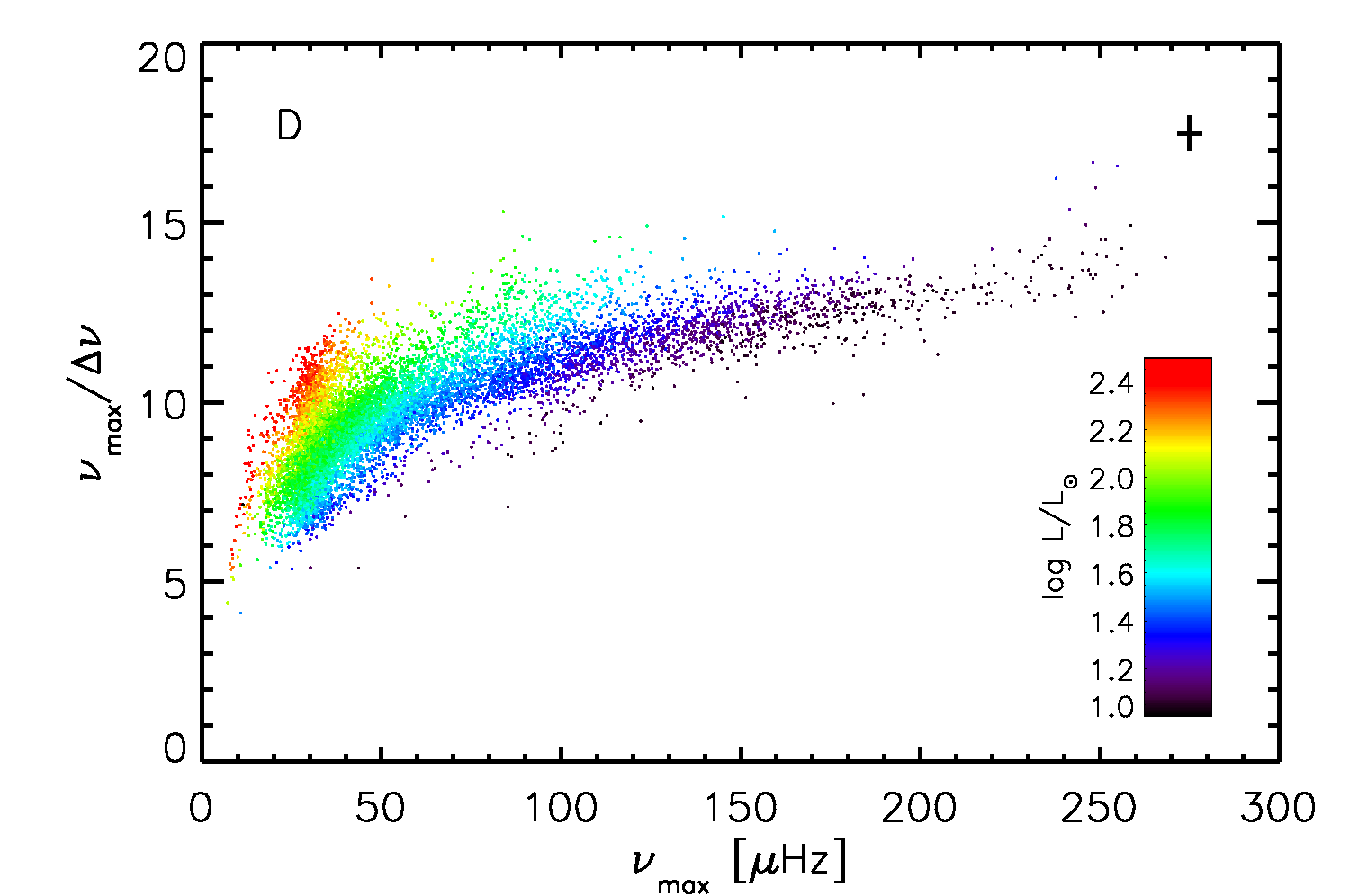}
\end{minipage}
\begin{minipage}{0.5\linewidth}
\centering
\includegraphics[width=\linewidth]{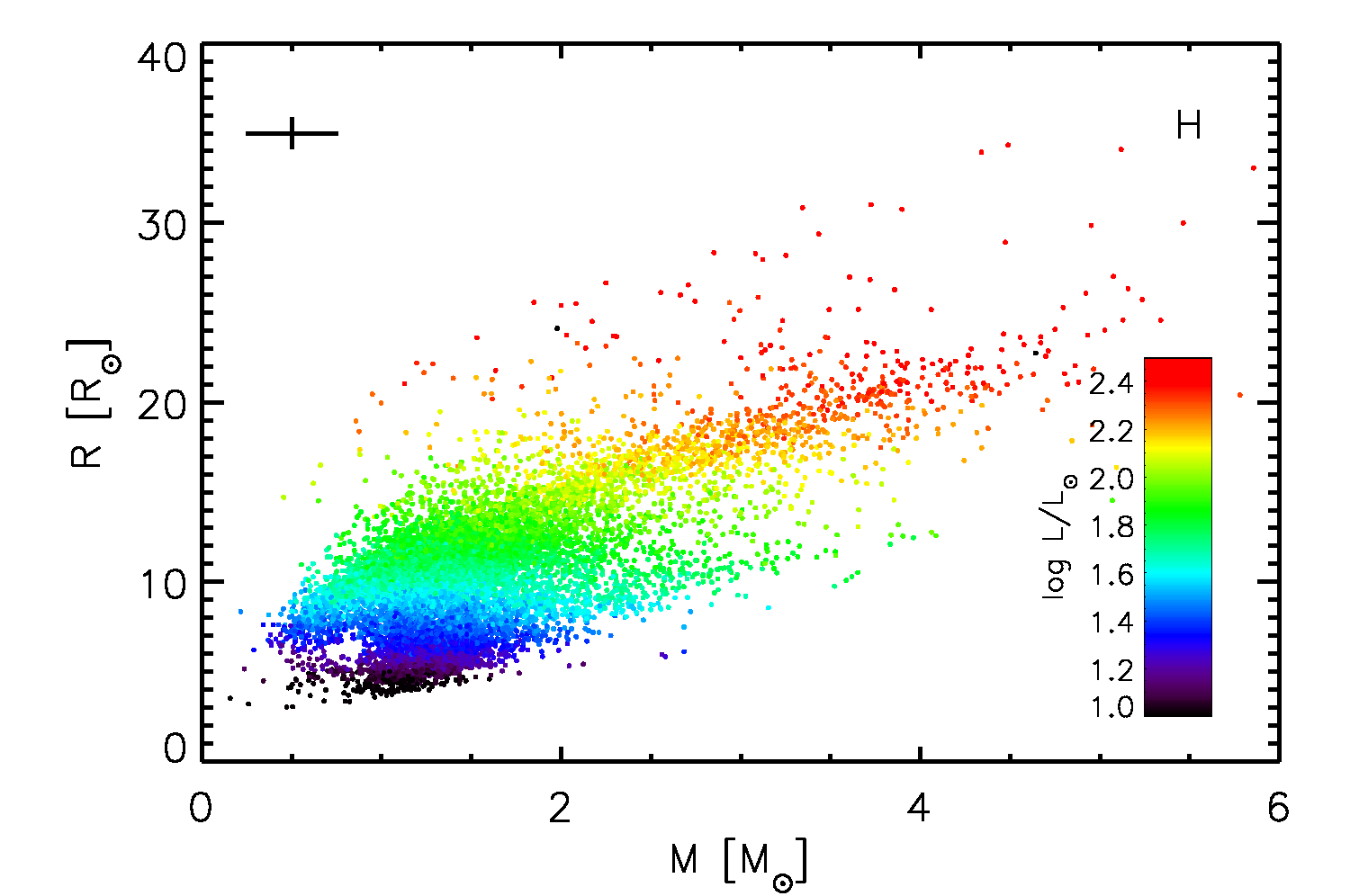}
\end{minipage}
\caption{$\nu_{\rm max}$ versus $\nu_{\rm max}/\Delta \nu$ diagrams (left) and H-R diagrams and mass versus radius diagrams (right) of the field red giants with detected oscillations. The colour-coding indicates from top to bottom mass, effective temperature, radius and luminosity. The highest and lowest values in the colour scale are lower and upper limits, respectively, to enhance the colour contrast. Characteristic uncertainties are indicated with a cross in the corner of each panel. The black dashed line in panel G indicates a radius of 7.5~R$_{\odot}$. Figure taken from \citet{hekker2011pub}.}
\label{res}
\end{figure*}

From the nearly 12\,000 red giants for which oscillations have been detected some specific stages of stellar evolution could be identified as shown in the diagrams in Fig.~\ref{res} \citep[see also][]{hekker2011pub}: 
\begin{itemize}
\item The majority of stars with 25~$\mu$Hz~$<$~$\nu_{\rm max}$~$<$~45~$\mu$Hz with masses roughly below 2~M$_{\odot}$ are known to be He-core burning stars that have gone through the Helium flash. These stars have radii of $\sim$10~R$_{\odot}$ and $\log L/\rm L_{\odot}$ of $\sim$2 and are the so-called \textit{red-clump stars} \citep[see e.g.,][for previous detections of the red clump]{miglio2009,huber2010,kallinger2010kepler,mosser2011}. 
\item The majority of stars with radii below $\sim$7.5~R$_{\odot}$ have low luminosities and lie  below the black dashed line in the H-R diagram (panel G in Fig.~\ref{res}). These are \textit{low-luminosity red-giant branch stars}. 
\item At $\nu_{\rm max}$~$<$~25~$\mu$Hz a group of stars with low effective temperatures is present (see panel B in Fig.~\ref{res}). These stars also have relatively high luminosities (see panel D) and large radii (see panel C), and low to intermediate masses (see panel A). These stars could be \textit{high-luminosity red-giant branch stars} or \textit{asymptotic giant branch stars}.
\item There is some evidence for a secondary clump. Stars with masses roughly between 2 and 3~M$_{\odot}$ and radii between 7.5 and 10~R$_{\odot}$ seem to form a separate branch in the mass versus radius diagram with slightly increased temperatures (panel F in Fig.~\ref{res}). These stars are also visible in the $\nu_{\rm max}$ versus $\nu_{\rm max}/\Delta\nu$ diagram at the expected location of the secondary clump \citep{girardi1999,huber2010,kallinger2010kepler}. These are stars in their He-burning phase, which are massive enough to have ignited He-burning in a non-degenerate core. These stars are \textit{secondary-clump stars}.
\item The high-mass stars ($\geq$~3.5~M$_{\odot}$) form a distinct group of stars at high $\nu_{\rm max}/\Delta\nu$ and $\nu_{\rm max}$ values similar to the red-clump stars (see panel A of Fig.~\ref{res}). These stars also have large radii and high luminosities.
\end{itemize}

\subsection{Clusters}
\begin{figure}
\centering
\begin{minipage}{0.6\linewidth}
\centering
\includegraphics[width=\linewidth]{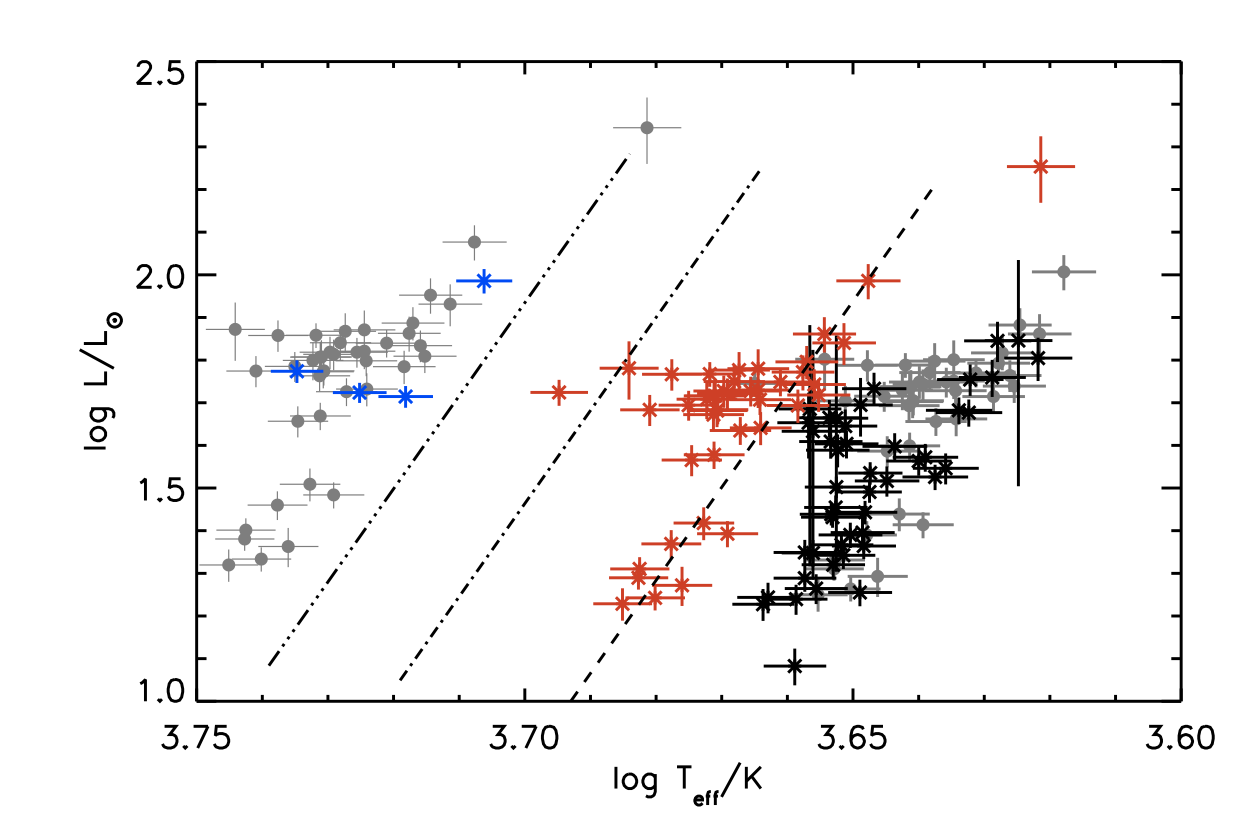}
\end{minipage}
\caption{H-R diagram of the clusters with NGC~6791 in black, NGC~6819 in red and NGC~6811 in blue. The gray symbols indicate NGC~6819 shifted to the positions of NGC~6791 and NGC~6811 (metallicity of $-$0.7~dex), respectively. The dashed line roughly indicates the red-giant branch of NGC~6819. The dashed-dotted  line indicates the position of the red-giant branch of NGC~6819 when only the mass difference with NGC~6811 is taken into account. The dashed-dotted-dotted-dotted line illustrates the position of the red-giant branch of NGC~6819 when the mass difference with NGC~6811is considered and a metallicity of $-$0.35~dex for NGC~6811 is assumed. Figure taken from \citet{hekker2011clus}.}
\label{HRclusters}
\end{figure}

From the global asteroseismic parameters and derived stellar parameters (luminosity, effective temperature, mass, and radius) of red giants in the three open clusters NGC~6791, NGC~6819 and NGC~6811 the following could be concluded \citep[see also][]{hekker2011clus}:
\begin{itemize}
\item Mass has a significant influence on the $\Delta \nu$ - $\nu_{\rm max}$ relation, while the influence of metallicity is negligible, under the assumption that the metallicity does not influence the excitation / damping of the oscillations. This has been predicted from models (A. Miglio and T. Kallinger, private communication), but now also clearly shown in observed data.
\item It is well known that both mass and metallicity have an influence on the position of stars in the H-R diagram. The different positions of the old metal-rich cluster NGC~6791 and the middle-aged solar-metallicity cluster NGC~6819 can indeed be explained by the observed differences in metallicity and mass.  For NGC~6811, with currently unknown metallicity, a metallicity of about $-$0.35 dex is needed to explain the position if the observed stars are He-core burning red-clump stars. However, if these stars are H-shell burning stars ascending the red-giant branch, the location of the stars in the H-R diagram can only be explained when the cluster has a subsolar metallicity of about $-$0.7 dex, see Fig.~\ref{HRclusters}.
\end{itemize}

\section{Discussion}
Results from ensemble asteroseismology as discussed here and in various other works \citep[e.g.,][]{deridder2009,miglio2009,chaplin2011,bedding2011nature} have shown to be of prime importance for asteroseismology. Stars have been observed in many varieties in terms of their stellar parameters, such as metallicity, rotation, mixing, magnetic fields etc. and in different evolutionary phases. Often, it remains difficult to determine whether an individual star under investigation is a `typical' example or a special case. In an ensemble study the few special cases will not get a lot of attention, but it becomes possible to investigate the behaviour of a particular group of stars with similar properties. Furthermore, results of in-depth studies of individual stars are in many cases model dependent. This model dependence does not disappear by a homogeneous study of an ensemble, but when comparing different ensembles one can draw conclusions based on the relative values of parameters.

Some recent examples of breakthrough results from ensemble asteroseismology of red-giant stars are the unambiguous detection non-radial oscillations \citep{deridder2009}, which opened the possibility to study the internal structures of these stars. However, stars in the H-shell burning phase (ascending the red-giant branch) and in the He-core burning phase (red clump) can have similar masses and radii, and it remained difficult to identify the evolutionary state of a particular star. The location in the H-R~diagram for different evolutionary states determined from a statistically significant ensemble of stars as summarised in this article increased the probability with which the evolutionary state of a particular star could be determined. More recently, oscillation modes that are sensitive to the core have been detected \citep{beck2011} and ensemble studies have shown that the period spacing between these modes for stars ascending the red-giant branch and in the red clump are significantly different \citep{bedding2011nature,mosser2011dp}. This is a major breakthrough and an excellent example of the benefits of ensemble asteroseismology: the meaning of the relative values of the period spacings became clear in an ensemble study.

The general behaviour of a particular ensemble can also be used to investigate other parameters. The investigation of the metallicity of NGC~6811 with respect to the metallicity of NGC~6819 as discussed here is an example of that. Other studies have focussed on for instance population studies \citep[e.g.,][]{miglio2009,chaplin2011}. \citet{miglio2009} showed that based on an ensemble of stars observed with CoRoT a constant star formation rate in our galaxy is more likely than a recent star formation burst. This is supported by a similar study of main-sequence stars observed with \textit{Kepler} \citep{chaplin2011}.

Ensemble asteroseismology or ensemble studies in general can be used in a wide variety of applications, such as to distinguish between for instance stars in the thin disk, thick disk or the bulge of our galaxy, or between population I and population II stars, evolution of stars with or without convective cores in their main-sequence phase etc. These ensemble studies are of prime importance and produce essential and complementary results compared to in-depth studies of individual stars. \newline 

\acknowledgements These proceedings are dedicated to the victims of the tragic disaster of the earthquake and tsunami that occurred in Japan on March 11, 2011 and following days. Despite these difficult circumstances, the Fujihara seminar: `Progress in solar/stellar physics with helio- and asteroseismology' has been a fruitful meeting, thanks to the excellent organisation. The authors thank the entire \textit{Kepler} team, without whom these results would not have been possible. SH acknowledges financial support from the Netherlands Organisation for Scientific Research (NWO), Leidsch Kerkhoven-Bosscha Fonds (LKBF) and the Fujihara foundation.

\bibliographystyle{asp2010}
\bibliography{hekker}

\end{document}